\documentclass[11pt]{article} 
\usepackage{graphicx,amssymb,mathrsfs,array,multirow}  
\begin{document}  
 
\vskip 30pt

\begin{center}  
{\Large{\bf 
Radiative generation of realistic neutrino mixing with $A4$ }}\\
\vspace*{1cm}  
\renewcommand{\thefootnote}{\fnsymbol{footnote}}  
{ {\sf Soumita Pramanick$^{1}$\footnote{email: soumita509@gmail.com}}
} \\  
\vspace{10pt}  
{\small  {\em $^1$Harish-Chandra Research Institute, Chhatnag Road, Jhunsi,
Allahabad 211019, India \\
}}
\normalsize  

\normalsize

\end{center}  

\begin{abstract} 
\textit{
Radiative generation of realistic mixing in neutrino sector
is studied at one-loop
level in a
scotogenic $A4\times Z_2$ symmetric framework.
A scheme of obtaining  non-zero $\theta_{13}$
through small mass splitting in right-handed neutrino
sector is proposed. 
The model consists of three right-handed neutrinos,
two of which were required to be degenerate in masses
to yield the common structure of the left-handed neutrino mass 
matrix that corresponds to $\theta_{13}=0$, $\theta_{23}=\pi/4$ and any $\theta_{12}^0$ 
in particular the choices specific to the Tribimaximal (TBM),
Bimaximal (BM) and Golden Ratio (GR) mixings.
Non-zero $\theta_{13}$, deviations of $\theta_{23}$ from
maximality and small corrections to the solar mixing angle
$\theta_{12}$ can be generated in one stroke by shifting from this 
degeneracy in the right-handed neutrino sector by a small amount.
The lightest among the three $Z_2$ odd inert $SU(2)_L$ doublet scalars 
present in the model can be a potential dark matter candidate.
}

\end{abstract}  

\renewcommand{\thesection}{\Roman{section}} 
\setcounter{footnote}{0} 
\renewcommand{\thefootnote}{\arabic{footnote}} 
\noindent


\section{Introduction}
Neutrino oscillation observations have clearly demonstrated the massive
nature of neutrinos. The mass eigenstates are 
 non-degenerate and distinct from the flavour
eigenstates and are connected to each other by the Pontecorvo, Maki, Nakagawa, Sakata --
PMNS -- matrix usually parametrized as:
\begin{eqnarray}
U = \left(
          \begin{array}{ccc}
          c_{12}c_{13} & s_{12}c_{13} & -s_{13}e^{-i\delta}  \\
 -c_{23}s_{12} + s_{23}s_{13}c_{12}e^{i\delta} & c_{23}c_{12} +
s_{23}s_{13}s_{12}e^{i\delta}&  -s_{23}c_{13}\\
 - s_{23}s_{12} - c_{23}s_{13}c_{12}e^{i\delta}& - s_{23}c_{12} +
c_{23}s_{13}s_{12}e^{i\delta} & c_{23}c_{13} \end{array} \right)
\;\;,
\label{PMNS}
\end{eqnarray}
where $c_{ij} = \cos \theta_{ij}$ and $s_{ij} = \sin \theta_{ij}$. 

In 2012, the short-baseline reactor anti-neutrino experiments observed
non-zero $\theta_{13}$, yet small compared to the other two mixing angles \cite{t13_s3}.
Prior to this observation, models leading to 
several structures like the Tribimaximal (TBM),
Bimaximal (BM) and Golden Ratio (GR) mixings (which we refer henceforth 
as popular lepton mixings) were studied all of which were
constructed with $\theta_{13}=0$ and $\theta_{23}=\pi/4$ and 
varying $\theta_{12}^0$ yielded the different
alternatives.

Putting $\theta_{13}=0$ and $\theta_{23}=\pi/4$ in
Eq. (\ref{PMNS}) can lead to the common structure for all popular mixings:
\begin{equation}
U^0=
\pmatrix{\cos \theta_{12}^0 & \sin \theta_{12}^0  & 0 \cr -\frac{\sin
\theta_{12}^0}{\sqrt{2}} & \frac{\cos \theta_{12}^0}{\sqrt{2}} &
-{1\over\sqrt{2}} \cr
-\frac{\sin \theta_{12}^0}{\sqrt{2}} & \frac{\cos
\theta_{12}^0}{\sqrt{2}}  & {1\over\sqrt{2}}},
\label{mix0}
\end{equation}
where $\theta_{12}^0$ for TBM, BM and GR are listed in Table \ref{t1}.

The present 3$\sigma$ global fits for the three mixing angles
\cite{Gonzalez_s3, Valle_s3}:
\begin{eqnarray}
\theta_{12} &=&(31.42 - 36.05)^\circ, \nonumber \\
\theta_{23} &=& (40.3-51.5)^\circ \,, \nonumber \\
\theta_{13} &=& (8.09 - 8.98)^\circ.
\label{results}
\end{eqnarray}
The numbers are from NuFIT3.2 of 2018 \cite{Gonzalez_s3}.

\begin{table}[tb]
\begin{center}
\begin{tabular}{|c|c|c|c|}
\hline
  Model &TBM &BM & GR \\ \hline
$\theta^0_{12}$ & 35.3$^\circ$ & 45.0$^\circ$  & 31.7$^\circ$ \\ \hline
\end{tabular}
\end{center}
\caption{\sf{$\theta^0_{12}$ for the different popular lepton mixings
 viz. TBM, BM, and GR  mixing.}}
\label{t1}
\end{table}

Thus the popular mixings are in disagreement with the observed non-zero $\theta_{13}$.
A plethora of activities had been taking place since this discovery 
to incorporate non-zero $\theta_{13}$ in these mixings.
Attempts to relate the smallness of solar splitting with that of $\theta_{13}$
can be found in \cite{br}.
In \cite{pr}, $\Delta
m^2_{atmos}$ and $\theta_{23} = \pi/4$ were embedded in the dominant
component of neutrino masses and mixing and the other oscillation parameters
such as $\theta_{13}, \theta_{12}$, the deviation
of $\theta_{23}$ from $\pi/4$, and $\Delta m^2_{solar}$
were obtained perturbatively from a smaller see-saw \cite{seesaw}
contribution\footnote{Some such earlier models can be found in 
\cite{old}.}. Vanishing $\theta_{13}$ can be induced by certain symmetries
and models generating non-zero $\theta_{13}$ through perturbation to 
such symmetric structures are also studied \cite{models, LuhnKing}.

In \cite{ourS3,newA4}, discrete flavour symmetries like $A4$, $S3$
were used to devise a two-component Lagrangian formalism at tree-level
to ameliorate all the popular mixing patterns in single stroke. 
The dominant contribution to the Lagrangian was obtained from Type II see-saw
mechanism characterized by popular mixing patterns, to which corrections were
obtained from a sub-dominant Type I see-saw contribution. 
In \cite{ourA4} the same scheme was performed for the no solar mixing (NSM) 
case i.e., $\theta_{12}^0=0$ case with $A4$ symmetry\footnote{ The dominant Type II see-saw
structure was kept devoid of solar splitting and degenerate perturbation theory was used 
to obtain large $\theta_{12}$.}. The basic difference between \cite{newA4,ourA4} and earlier 
works with $A4$ \cite{A4mr, A4af, otherA4} is that in the earlier works  Type II see-saw was used  to generate
the mass matrices and obtaining TBM was the prime goal. More realistic mixings can be found in
recent works \cite{newMa, tanimoto}.

In this paper, we intend to generate:
\begin{enumerate}
\item the popular mixing structure in Eq. (\ref{mix0}) 
with $\theta_{13}=0$, $\theta_{23}=\pi/4$ and $\theta_{12}^0$ as listed in Table \ref{t1} 
\item  non-zero $\theta_{13}$, deviations of $\theta_{23}$ from $\pi/4$ and
small corrections to $\theta_{12}$ 
\end{enumerate}
radiatively\footnote{For
review of radiative neutrino mass models see \cite{radreview}.} with $A4$ flavour symmetry\footnote{For a 
brief discussion on $A4$ see Appendix of the paper.}.
Precisely, this $A4\times Z_2$ symmetric model
will produce neutrino masses at one-loop level 
using three right-handed
neutrinos that transform as a triplet 
under $A4$. To get Eq. (\ref{mix0}) it is necessary that
two of these right-handed neutrinos are degenerate.
A little shift from that degeneracy will yield non-zero $\theta_{13}$,
deviations of atmospheric mixing from maximality and tinker
the solar mixing by a small amount in one go.

In order to accomplish this we also need to introduce a $Z_2$ odd
$A4$ triplet scalar field $\eta$, the lightest of which could be a potential
dark matter candidate.

\section{The Model}
The neutrino mass matrix in the mass basis is given by 
$M^{mass}_{\nu L}=$ $diag\; (m_1, m_2, m_3)$. This when expressed in 
flavour basis using the common structure of $U^0$ in Eq. (\ref{mix0}) for the popular
lepton mixings, leads to:
\begin{equation}
M^{flavour}_{\nu L}=U^0 M^{mass}_{\nu L} U^{0T}=\pmatrix{a & c & c \cr
c & b & d \cr c & d & b}
\label{abc}
\end{equation}
where,
\begin{eqnarray}
a&=& m_1\cos^2 \theta_{12}^0+m_2\sin^2 \theta_{12}^0\nonumber\\
b&=&\frac{1}{2}\left(m_1\sin^2 \theta_{12}^0+m_2\cos^2 \theta_{12}^0+m_3\right)\nonumber\\
c&=&\frac{1}{2\sqrt{2}}\sin 2\theta_{12}^0 (m_2-m_1)\nonumber\\
d&=&\frac{1}{2}\left(m_1\sin^2 \theta_{12}^0+m_2\cos^2 \theta_{12}^0-m_3\right)
\label{abcd}
\end{eqnarray}
Equivalently,
\begin{equation}
\tan 2\theta_{12}^0=\frac{2\sqrt 2 c}{b+d-a}
\end{equation}
For non-degenerate realistic neutrino masses
$a,b,c$ and $d$ are non-zero. 

Our objective is to obtain the structure of the matrix shown in Eq. (\ref{abc})
at one-loop level. For that we assign specific $A4\times Z_2$ charges to the scalars 
and fermions in our model. This model has three right-handed neutrino fields.
As we will see in course of the discussion that in order to obtain the structure in Eq. (\ref{abc}), 
two of these right-handed neutrino states will require to be degenerate in masses.
Once the structure in Eq. (\ref{abc}) is produced, we will exploit small relaxation of this degenerate
feature in right-handed neutrino sector to yield the realistic neutrino mixings, namely non-zero $\theta_{13}$. 

In this model, apart from the three $SU(2)_L$ lepton doublets
we have three right-handed neutrinos, $N_{\alpha R}$,
($\alpha=1,2,3$) invariant under the standard model (SM) gauge group. 
Under $A4$ these three right-handed neutrinos transform as a triplet
and so does the three $SU(2)_L$ lepton doublets. In the scalar sector 
we have two $A4$ symmetric triplet fields $\Phi$ and $\eta$ each of which
comprises of three $SU(2)_L$ doublet 
fields $\Phi_i\equiv(\phi_i^+, \phi_i^0)^T$ and $\eta_j\equiv(\eta_j^+, \eta_j^0)^T$, $(i,j=1,2,3)$.
In addition to $A4$ we have an unbroken $Z_2$ under which all the fields are even except
the scalar field $\eta$ and the right-handed neutrinos. Thus the scalars $\eta_j$ do not
acquire vacuum expectation value (vev) after spontaneous symmetry breaking (SSB),
whereas the fields $\phi_i$ do.
All the fields along with their quantum numbers are listed in Table \ref{fields}.
Here we restrict ourselves to the neutrino sector only\footnote{This model differs from \cite{Ma_rad} in terms of the particle content.
Unlike \cite{Ma_rad}, here we consider
all the popular mixings viz. TBM, BM and GR and also
generate non-zero $\theta_{13}$, deviations
of $\theta_{23}$ from maximality
and small corrections to $\theta_{12}$ simultaneously
through small mass splitting in the right-handed neutrino sector. }.
We work in a basis in which the charged lepton mass
matrix is diagonal and the mixing is entirely from the
neutrino sector. 

\begin{figure}[tbh]
\begin{center}
\includegraphics[scale=0.19,angle=0]{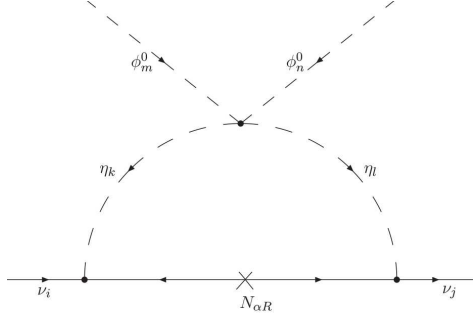}
\end{center}
\caption{\sf  Neutrino mass generation at one-loop level.
}
\label{radfig} 
\end{figure}

\begin{table}[tb]
\begin{center}
\begin{tabular}{|c|c|c|c|}
\hline
\sf{Leptons} & $SU(2)_L$ & $A4$ & $Z_2$ \\ \hline
 & & &  \\ 
$L_\beta\equiv\pmatrix{\nu_e& e^-\cr
\nu_\mu & \mu^- \cr \nu_\tau & \tau^- }$ & $2$ & $3$ & $1$ \\
 & & &  \\ 
 \hline
 & & &  \\ 
$N_{\alpha R}\equiv \pmatrix{N_{1R}\cr
N_{2R}\cr N_{3R}} $ & $1$ & $3$ & $-1$ \\ 
 & & &  \\ 
\hline
\hline
\sf{Scalars}& $SU(2)_L$ & $A4$ & $Z_2$ \\ \hline
 & & &  \\ 
$\Phi \equiv \pmatrix{\phi_1^+ & \phi_1^0\cr
\phi_2^+ & \phi_2^0 \cr \phi_3^+ & \phi_3^0 }$ & $2$ & $3$ & $1$ \\ 
 & & &  \\ 
\hline
 & & &  \\ 
$\eta\equiv \pmatrix{\eta_1^+ & \eta_1^0\cr
\eta_2^+ & \eta_2^0 \cr \eta_3^+ & \eta_3^0 }$ & $2$ & $3$ & $-1$ \\ 
 & & &  \\ 
\hline
\end{tabular}
\end{center}
\caption{\sf{Fields and their quantum numbers. Here we
are concerned with the neutrino sector only.}}
\label{fields}
\end{table}
With these fields one can generate neutrino mass at one-loop
level as shown in Fig. (\ref{radfig}).
The relevant part of the scalar potential from the four-point scalar vertex
contributing to the neutrino mass matrix is given by\footnote{Note at the
four-point scalar vertex, both the $\phi$ fields are annihilated and both the
$\eta$ fields are created. So terms of the scalar potential of 
$(\eta^\dagger \phi)(\eta^\dagger \phi)$ kind will contribute to the neutrino mass
matrix and are therefore relevant. }:

\begin{eqnarray}
V_{relevant}&\supset& \lambda_1 \left[ \left\{(\eta_1^\dagger \phi_1+\eta_2^\dagger \phi_2+ 
\eta_3^\dagger \phi_3)^2 \right  \}  + h.c.\right]\nonumber\\
&+&\lambda_2\left[ \left\{(\eta_1^\dagger \phi_1+ \omega\eta_2^\dagger \phi_2+ 
\omega^2\eta_3^\dagger \phi_3)(\eta_1^\dagger \phi_1+ \omega^2\eta_2^\dagger \phi_2+ 
\omega\eta_3^\dagger \phi_3) \right  \} +h.c.\right]\nonumber\\
&+&\lambda_3\left[ \left\{(\eta_2^\dagger \phi_3)^2+ (\eta_3^\dagger \phi_2)^2+ (\eta_3^\dagger \phi_1)^2
+(\eta_1^\dagger \phi_3)^2+ (\eta_1^\dagger \phi_2)^2+(\eta_2^\dagger \phi_1)^2 \right  \} 
+h.c.\right]\nonumber\\
&+&\lambda_4\left[ \left\{(\eta_2^\dagger \phi_3)(\eta_3^\dagger \phi_2)+ (\eta_3^\dagger \phi_1)
(\eta_1^\dagger \phi_3)+ (\eta_1^\dagger \phi_2)(\eta_2^\dagger \phi_1) \right  \} 
+h.c.\right],
\label{potential}
\end{eqnarray}
where all the quartic couplings $\lambda_i$ ($i=1,2,3,4$) are considered to be real.

As discussed earlier, after SSB, $\phi_i^0$ will get vevs whereas the $\eta_j^0$ will
not owing to the $Z_2$ assignments. Let $\langle\Phi_i\rangle=v_i$ where ($i=1,2,3$).
In \cite{gmin1, gmin2}, it has been shown that for $A4$ symmetric three-Higgs-doublets
the four vev configurations for which the scalar potential acquires the 
global minima\footnote{In \cite{our3hdm}, it has been shown that alignment follows as a natural
consequence of the discrete symmetry $A4$ in $A4$ symmetric three-Higgs-doublet model
for all the four global minima configurations mentioned in Eq. (\ref{vevs}).} are:
\begin{equation}
\langle \Phi \rangle_{case1} = v\pmatrix{0 & 1 \cr 0
& 0 \cr 0 & 0} \;,\;
\langle \Phi \rangle_{case2} = v\pmatrix{0 & 1 \cr 0
& e^{i \alpha} \cr 0 & 0} \;,\;
\langle \Phi \rangle_{case3} = v\pmatrix{0 & 1 \cr 0
& 1 \cr 0 & 1} \;,\;
\langle \Phi \rangle_{case4}= v \pmatrix{0 & 1 \cr 0
& \omega \cr 0 & \omega^2} \;.
\label{vevs}
\end{equation}

Let $\eta_{Rj}$ and $\eta_{Ij}$ be the real and
imaginary parts of $\eta_j^0$ respectively. Splitting among the
masses of $\eta_{Rj}$ and $\eta_{Ij}$ is proportional to $\lambda v_j$
and is expected to be small. Here
$\lambda$ stands for the quartic couplings 
$\lambda_1$, $\lambda_2$, $\lambda_3$ and
$\lambda_4$ in Eq. (\ref{potential}).
Also the mass splittings between $\eta_j$ ($j=1,2,3$) constituting the
$A4$ triplet are neglected and $m_0$ is their common mass.
Our model has three right-handed neutrinos, $N_{\alpha R}$ ($\alpha=1,2,3$),
$M_\alpha$ being their masses. In the limit $M_\alpha^2>>m_0^2$,
the diagram in Fig. (\ref{radfig}) leads to neutrino mass of the kind \cite{Ma_loop}:
\begin{equation}
(M_{\nu_L}^{flavour})_{ij} = \lambda \frac{v_m v_n}{8 \pi^2}
\sum_{\alpha,k,l\ne (i,j)} 
\frac{h_{i\alpha k} h_{j\alpha l}}{M_\alpha} 
\left[ \ln z_\alpha -1 \right].
\label{loop}
\end{equation}
where $z_\alpha\equiv\frac{M_\alpha^2}{m_0^2}$. 
The vevs of $\phi^0_m$ and $\phi^0_n$ are given by $v_m$ and $v_n$
respectively. 
The Yukawa couplings at the two vertices where the left-handed
and right-handed neutrinos couple to the inert $SU(2)_L$ doublet fields
$\eta$ are given by $h_{i\alpha k}$ and $h_{j\alpha l}$.
No two of the three indices appearing in each of $h_{i\alpha k}$ and $h_{j\alpha l}$
individually can be same owing to $A4$ symmetry.
Thus $h_{i\alpha k}$ and $h_{j\alpha l}$ are determined by the
$A4$ invariance which in its turn governs the structure of the neutrino
mass matrix. 
Since the logarithm is a slowly varying
function and the heavy right handed neutrino masses
$M_\alpha$ ($\alpha=1,2,3$) are expected to be close
to each other, the RHS of Eq. (\ref{loop}) can be approximated
as proportional to $\frac{1}{M_\alpha}$.
Leaving the vevs $v_m$, $v_n$ and the quartic couplings
$\lambda$, let us denote the contribution to left-handed
neutrino mass matrix $(M_{\nu_L}^{flavour})_{ij}$ from everything 
else in Eq. (\ref{loop}) by loop contributing factors $r_\alpha\propto \frac{1}{M_\alpha}$.

For simplicity let us consider the right-handed neutrino mass matrix to be
already diagonal i.e., $M_{N_R}\equiv diag (M_1, M_2, M_3)$.
In terms of the right-handed neutrino loop contributing factors $r_\alpha$
we have the contribution coming from right-handed neutrino sector as
$diag(r_1, r_2, r_3)$. Using Eqs. (\ref{loop}) and (\ref{potential}), the
left-handed neutrino mass matrix that arises from Fig. (\ref{radfig}) is given by\footnote{All
the symmetries under consideration were conserved at each of the three vertices in Fig. (\ref{radfig}).}:

\begin{equation}
M_{\nu_L}^{flavour}=\pmatrix{ \chi_1
& \chi_4 & \chi_5 \cr
\chi_4
& \chi_2 & \chi_6 \cr 
\chi_5
& \chi_6 & \chi_3
}
\label{mgeneral}
\end{equation}
where,
\begin{eqnarray}
\chi_1&\equiv&(\lambda_1+\lambda_2)(r_3v_2^2+r_2v_3^2)
+\lambda_3[r_2(v_1^2+v_2^2)+r_3(v_1^2+v_3^2)]\nonumber\\
\chi_2&\equiv&(\lambda_1+\lambda_2)(r_1v_3^2+r_3v_1^2)
+\lambda_3[r_1(v_1^2+v_2^2)+r_3(v_2^2+v_3^2)]\nonumber\\
\chi_3&\equiv&(\lambda_1+\lambda_2)(r_1v_2^2+r_2v_1^2)
+\lambda_3[r_2(v_2^2+v_3^2)+r_1(v_1^2+v_3^2)]\nonumber\\
\chi_4&\equiv& r_3 [\lambda_4+2\lambda_1-\lambda_2] v_1v_2\nonumber\\
\chi_5&\equiv& r_2 [\lambda_4+2\lambda_1-\lambda_2] v_1v_3\nonumber\\
\chi_6&\equiv& r_1 [\lambda_4+2\lambda_1-\lambda_2] v_2v_3.
\end{eqnarray}

In order to obtain the neutrino mass matrix of the form of 
Eq. (\ref{abc}) from Eq. (\ref{mgeneral}), one will simultaneously require
$\chi_1\ne\chi_2=\chi_3$ and $\chi_4=\chi_5$.
Let us now try each of the vev configurations in Eq. (\ref{vevs}) and
find out the one suitable to obtain this feature along with the constraints put on 
to $r_1, \, r_2$  and $r_3$.

\begin{enumerate}
\item Choice A: For $(v_1,v_2,v_3)=v(1,0,0)$, irrespective of the choices for 
 $r_1, \, r_2$  and $r_3$, the off-diagonal entries in Eq. (\ref{mgeneral}) will vanish
and one cannot obtain mixing in the neutrino sector.
\item Choice B: For $(v_1,v_2,v_3)=v(1,e^{i\alpha},0)$, two of the three off-diagonal entries
in  Eq. (\ref{mgeneral}) will vanish for any $r_1, \, r_2$  and $r_3$, and one cannot obtain the 
structure in Eq. (\ref{abc}).
\item Choice C: For $(v_1,v_2,v_3)=v(1, \omega, \omega^2)$, one cannot achieve
$\chi_2=\chi_3$ as required to obtain the structure of the mass matrix in
Eq. (\ref{abc}) from Eq. (\ref{mgeneral}), whatever may be the choices for $r_1,\,r_2,\, r_3$.
\item Choice D: For $(v_1,v_2,v_3)=v(1,1,1)$, note first that $r_1=r_2=r_3$
implies all the diagonal terms to be equal to each other and the off-diagonal
entries are equal among themselves. This leads to two left-handed degenerate states
and only TBM is admissible. We will not consider that choice. However the form in Eq. (\ref{abc})
starting from Eq. (\ref{mgeneral}) is achieved for $r_1\ne r_2=r_3=r$ when 
$(v_1,v_2,v_3)=v(1,1,1)$, which we refer to as choice D from now onwards.
This choice allows all three mixings viz. TBM, BM, GR and all three left-handed neutrinos
to be non-degenerate. 
Hence we will consider this case for
further analysis.
Such choice of $r_2=r_3$ is achieved when  
the right-handed neutrinos $N_{2R}$ and $N_{3R}$ are degenerate in masses.
\end{enumerate}
Putting choice D ie., $v_1=v_2=v_3=v$ and $r_1\ne r_2=r_3=r$ in 
Eq. (\ref{mgeneral}) one gets the following form of the left-handed 
neutrino mass matrix in the flavour basis:
\begin{equation}
M_{\nu_L}^{flavour}=\pmatrix{ \lambda_{123} (2 rv^2)
& \lambda_{124} rv^2 & \lambda_{124} rv^2\cr
\lambda_{124} rv^2
& \lambda_{123}  (r+r_1) v^2 & \lambda_{124} r_1v^2\cr 
\lambda_{124} rv^2
& \lambda_{124} r_1v^2 & \lambda_{123}(r+r_1) v^2
}
\label{mchoiceD}
\end{equation}
where, $\lambda_{123}=\lambda_1+\lambda_2+2\lambda_3$ and
$\lambda_{124}=\lambda_4+2\lambda_1-\lambda_2$.
Thus the neutrino mass matrix generated at one-loop
level as shown in Fig. (\ref{radfig}) can produce the form of $M_{\nu_L}^{flavour}$ 
as in Eq. (\ref{abc}) that
corresponds to $\theta_{13}=0$, $\theta_{23}=\pi/4$ and $\theta_{12}^0$ of the
popular mixing alternatives, with the vevs and right-handed neutrino masses as specified 
in choice D. This follows from the identifications:
\begin{eqnarray}
a&\equiv&\lambda_{123} (2 r_2v^2)= (\lambda_1+\lambda_2+2\lambda_3)(2 r_2v^2)\nonumber\\
b&\equiv&\lambda_{123}  (r+r_1) v^2= (\lambda_1+\lambda_2+2\lambda_3)(r+r_1) v^2\nonumber\\
c&\equiv&\lambda_{124} rv^2=(\lambda_4+2\lambda_1-\lambda_2)rv^2\nonumber\\
d&\equiv&\lambda_{124} r_1v^2=(\lambda_4+2\lambda_1-\lambda_2)r_1v^2
\label{id1}
\end{eqnarray} 

Having achieved this, next we concentrate on generation of 
realistic neutrino mixing i.e., 
non-zero $\theta_{13}$, deviations of $\theta_{23}$ from maximality
and small corrections in the solar mixing $\theta_{12}$.
For that one has to deviate from the $r_\alpha$ ($\alpha =1,2,3$) of choice D.
Let us now split the degeneracy in the right handed neutrino sector by a small
amount $\epsilon$ i.e., consider $r_3=r_2+\epsilon$ and $r_1\ne r_2\ne r_3\ne r_1$, keeping the
vevs still to be $v_1=v_2=v_3=v$. With such a choice one is expected to get 
a dominant contribution of the form
of $M_{\nu_L}^{flavour}$ as was achieved in Eq. (\ref{mchoiceD}), say $M^0$,  together
with small shift from it, $M'$, proportional to $\epsilon$. 
Thus,
\begin{equation}
M_{\nu_L}^{flavour}=M^0+M'
\end{equation}   
where, 
\begin{equation}
M^0=\pmatrix{ \lambda_{123} (2 r_2v^2)
& \lambda_{124} r_2v^2 & \lambda_{124} r_2v^2\cr
\lambda_{124} r_2v^2
& \lambda_{123}  (r_1+r_2) v^2 & \lambda_{124} r_1v^2\cr 
\lambda_{124} r_2v^2
& \lambda_{124} r_1v^2 & \lambda_{123}(r_1+r_2) v^2
}\, {\rm and} \, 
M'=\epsilon\pmatrix{ x
& y & 0\cr
y
& x & 0\cr 
0
& 0 & 0
}
\end{equation} 
where $x=\lambda_{123} v^2$ and $y=\lambda_{124} v^2$.
Here $M^0$ is the form of the $M_{\nu_L}^{flavour}$ 
required for $\theta_{13}=0$, $\theta_{23}=\pi/4$ and $\theta_{12}^0$ of the
popular mixings. Thus in analogy to Eq. (\ref{id1}), one can 
identify the followings\footnote{To distinguish from $r_2=r_3$ case,
let us use a primed notation.}:
\begin{eqnarray}
a'&\equiv&\lambda_{123} (2 rv^2)= (\lambda_1+\lambda_2+2\lambda_3)(2 rv^2)\nonumber\\
b'&\equiv&\lambda_{123}  (r_1+r_2) v^2= (\lambda_1+\lambda_2+2\lambda_3)(r_1+r_2) v^2\nonumber\\
c'&\equiv&\lambda_{124} r_2v^2=(\lambda_4+2\lambda_1-\lambda_2)r_2v^2\nonumber\\
d'&\equiv&\lambda_{124} r_1v^2=(\lambda_4+2\lambda_1-\lambda_2)r_1v^2
\label{id2}
\end{eqnarray} 
It is straightforward to incorporate the corrections offered by $M'$ to $M^0$
using the non-degenerate perturbation theory. Columns of $U^0$ in Eq. (\ref{mix0}) is the unperturbed
flavour basis. From Eq. (\ref{id2})
one can define:
\begin{equation}
 \gamma\equiv(b'-3d'-a') \ \  {\rm and} \ \ 
\rho\equiv\sqrt{a^{'2}+b^{'2}+8c^{'2}+d^{'2}-2a^{'}b^{'}-2a^{'}d^{'}+2b^{'}d^{'}}
\label{grho}
\end{equation}
The first order corrected third ket is then given by:
\begin{equation}
|\psi_3\rangle =
\pmatrix{\frac{\epsilon}{\gamma^2-\rho^2}
\left[\gamma(x\sin 2\theta_{12}^0-\sqrt{2}y\cos 2\theta_{12}^0)
+\rho\sqrt{2}y\right]
\cr
-\frac{1}{\sqrt{2}}[1+\xi \epsilon]
\cr
\frac{1}{\sqrt{2}}[1-\xi \epsilon]}.
\label{ket3}
\end{equation}
where, 
\begin{equation}
\xi\equiv[\gamma x +\rho(x\cos 2\theta_{12}^0+\sqrt{2}y\sin 2\theta_{12}^0)]/(\gamma^2-\rho^2).
\label{xi}
\end{equation}
Thus one can write,\footnote{Here we restrict ourselves to no CP-violation.}:
\begin{equation}
\sin \theta_{13}=\frac{\epsilon}{\gamma^2-\rho^2}\left[
\gamma(x\sin 2\theta_{12}^0-\sqrt{2}y\cos 2\theta_{12}^0)
+\rho\sqrt{2}y
\right].
\label{s13}
\end{equation}
Using Eqs. (\ref{id2}), (\ref{grho}) and (\ref{s13}),
one can easily read off non-zero $\theta_{13}$ in terms of the
model parameters, namely, $\epsilon$, the quartic couplings and the vevs.
Throughout our discussion we have assumed $r_\alpha$ ($\alpha=1,2,3$)
are real and restricted ourselves to a CP-conserving scenario. 
In principle, the right-handed neutrino masses
can have Majorana phases causing these $r_\alpha$ to be
complex. Then one can have a complex $\epsilon$, from which 
one can generate CP-violation in the lepton sector.

From Eq.(\ref{ket3}) the deviation of atmospheric mixing 
from maximality is given by:
\begin{equation}
\tan\varphi\equiv\tan(\theta_{23}-\pi/4)=\xi \epsilon.
\label{atmmix}
\end{equation}
Similarly, one can obtain small corrections to $\theta_{12}$
from the corrections of the first and second kets. 
The solar mixing angle after receiving first
order corrections is given by:
\begin{equation}
\tan\theta_{12}=\frac{\sin\theta_{12}^0+\epsilon \beta \cos\theta_{12}^0}{\cos\theta_{12}^0-\epsilon \beta \sin\theta_{12}^0}
\label{solmix}
\end{equation}
where,
\begin{equation}
\beta\equiv \frac{\left[ \frac{y}{\sqrt{2}}\cos 2\theta_{12}^0
+\frac{x}{\sqrt{4}}\sin 2\theta_{12}^0
\right]}{\rho}
\label{beta}
\end{equation}
The corrections to the solar mixing and deviations
of atmospheric mixing from $\pi/4$ in 
Eq. (\ref{solmix}) and (\ref{atmmix}) respectively can be expressed
in terms of the model parameters using Eqs. (\ref{id2}), 
(\ref{grho}), (\ref{xi}) and (\ref{beta}). 

Summing up, a scotogenic $A4\times Z_2$ symmetric model
of radiatively obtaining realistic neutrino mixing is
proposed. 
Among others, the model comprises of three gauge singlet
right-handed neutrino fields $N_{\alpha R}$, ($\alpha=1,2,3$).
If $N_{2 R}$ and $N_{3 R}$ are degenerate in masses, one can obtain
the common structure of the left-handed neutrino mass matrix required by
$\theta_{13}=0$, $\theta_{23}=\pi/4$ and $\theta_{12}^0$ of the
particular choices leading to popular lepton mixing scenarios viz. TBM, BM, GR
at one-loop level. 
A slight shift from this degeneracy of right-handed neutrino masses could generate 
realistic mixing viz.
non-zero $\theta_{13}$, deviations of
$\theta_{23}$ from $\pi/4$
and also tweak $\theta_{12}$ by a small amount. The model has three inert $SU(2)_L$ doublet scalars
$\eta$, odd under the unbroken $Z_2$, the lightest of which can be a dark matter candidate.

{\bf Acknowledgements:} 
I thank
Prof. Amitava Raychaudhuri for discussions at
different stages of this work.

\renewcommand{\thesection}{\Alph{section}} 
\setcounter{section}{0} 
\renewcommand{\theequation}{\thesection.\arabic{equation}}

\setcounter{equation}{0}

\section{Appendix: The discrete group $A4$ }\label{GroupA4}
$A4$ being the group of even permutations of four objects has 
12 elements. The group $A4$ has two generators $S$ and $T$. These generators satisfy
$S^2=T^3=(ST)^3=\mathbb{I}$.
The inequivalent irreducible representations for $A4$ are four in number
out of which three are 1-dimensional viz. $1, 1'$ and
$1''$ and one is 3-dimensional. The 1-dimensional representations
transform as 1, $\omega$, and $\omega^2$ under\footnote{Here
$\omega$ is a cube root of 1.} $T$ but are invariant
under $S$. Thus, $1'\times 1'' = 1$. The generators are represented by,  
\begin{equation}
S=\pmatrix{1 & 0 & 0 \cr 0 & -1 & 0 \cr 0 & 0 & -1}\ \ \ \ {\rm and} \ \ \ \
T=
\pmatrix{0 & 1 & 0 \cr
0 & 0 & 1 \cr
1 & 0 & 0} \;\;.
\label{ST3}
\end{equation}
Below is the combination rule for two $A4$ triplets:
\begin{equation}
3\otimes 3=1 \oplus 1' \oplus 1'' \oplus 3 \oplus 3 \;\;.
\label{A43x3}
\end{equation}
Let us have two $A4$ triplet fields, $3_a \equiv {a_i}$ and $3_b \equiv {b_i}$, where
$i=1,2,3$, and combine them according to Eq. (\ref{A43x3}). 
The triplets that we
get can be written as $3_c \equiv {c_i}$
and $3_d \equiv {d_i}$ where, 
\begin{eqnarray}
c_i &=& \left(\frac{a_2 b_3 + a_3 b_2}{2},\frac{a_3 b_1 + a_1 b_3}{2},
\frac{a_1 b_2 + a_2 b_1}{2}\right)   \;\;, \;\;{\rm or,}
\;\; c_i \equiv \alpha_{ijk} a_j b_k \;\;, \nonumber \\
d_i &=& \left(\frac{a_2 b_3 - a_3 b_2}{2},\frac{a_3 b_1 - a_1 b_3}{2},
\frac{a_1 b_2 - a_2 b_1}{2}\right)   \;\;, \;\;{\rm or,}
\;\; d_i \equiv \beta_{ijk} a_j b_k \;\;,\;\; (i,j,k, {\rm are ~cyclic})\;\;.
\label{3x3to3}
\end{eqnarray}
The $1$, $1'$ and $1''$ in this case are: 
\begin{eqnarray}
1 &=& a_1b_1+a_2b_2+a_3b_3 \equiv \rho_{1ij}a_ib_j \;\;,\nonumber \\  
1'&=& a_1b_1+\omega^2a_2b_2+\omega a_3b_3 \equiv \rho_{3ij}a_ib_j
\;\;,  \nonumber \\
1'' &=& a_1b_1+\omega a_2b_2+\omega^2a_3b_3 \equiv \rho_{2ij}a_ib_j \;\;. 
\label{3x3to1}
\end{eqnarray}
The group was studied in context 
of neutrino mass and mixings in the pioneering works \cite{A4mr, A4af}.

\end{document}